\documentclass{hep99}

\usepackage{epsfig}
\usepackage{amssymb}
\usepackage{mcite}

\def\Journal#1#2#3#4{{#1} {\bf #2} (#4) #3}
\def\NIM{{\em Nucl. Instr. and Meth.} A}
\def\NPB{{\em Nucl. Phys.} B}
\def\PLB{{\em Phys. Lett.}  B}

\def\PRD{{\em Phys. Rev.} D}
\def\PRX{{\em Phys. Rev.}}
\def\ZPC{{\em Z. Phys.} C}
\def\ZPX{{\em Z. Phys.}}
\def\EPC{{\em Eur. Phys. Jour.} C}
\def\CPC{{\em Comp. Phys. Comm.}}

\def\gp{\gamma p}
\def\q2{Q^2}
\def\kt{k_T}
\def\pb1{pb$^{-1}$}
\def\etjet{E_T^{jet}}
\def\etajet{\eta^{jet}}
\def\phijet{\varphi^{jet}}
\def\g2{GeV$^2$}
\def\set{d\sigma/d\etjet}
\def\mj{M^{JJ}}
\def\cost{\vert\cos\theta^*\vert}

\def\scost{d\sigma/d\cost}

\def\costh3{\cos\theta_3}
\def\xo{x_{\gamma}^{OBS}}
\def\oaa{{\cal O}(\alpha\alpha_s^2)}
\def\rr1{R=1}

\begin{document}

\title{QCD Tests at HERA$^\dag$}

\author{C Glasman$^{a*}$\\ representing the H1 and ZEUS Collaborations}

\address{$^a$ Departamento de F\'\i sica Te\'orica,
Universidad Aut\'onoma de Madrid,\\
Cantoblanco, 28049 Madrid, Spain\\[3pt]
E-mail: {\tt claudia@mail.desy.de}}

\abstract{Measurements of inclusive jet, dijet and three-jet cross sections in
photon-proton interactions are presented. These measurements provide new
tests of QCD, constrain the parton densities of the photon, and allow
searches for new physics. Measurements of the mean subjet multiplicity are
reported and used to test the differences between quark and gluon jets.}

\maketitle

\fntext{\dag}{Talk given at the {\it International Europhysics
Conference on High Energy
Physics}, Tampere, Finland, July $15^{th}-21^{th}$, 1999.\\
$^*$ Supported by an EC fellowship number ERBFMBICT 972523.}

\section{Introduction}
Measurements of cross sections for processes involving a large momentum
transfer (e.g. jets in $\gp$ interactions) are compared to next-to-leading
order (NLO) calculations to test QCD. These measurements also provide a test
of the parametrisations of the photon parton densities
and may be used in global analyses to constrain the parton distributions.

The investigation of the internal structure of jets gives insight into the
transition between a parton produced in a hard process and the experimentally
observable spray of hadrons. Measurements of jet substructure allow the study
of the characteristics of quark and gluon jets.

\section{Jet cross sections in $\gp$ interactions}
\begin{figure*}
\begin{center}
\setlength{\unitlength}{1.0cm}
\begin{picture} (5.0,4.0)
\put (-3.0,0.2){\epsfig{figure=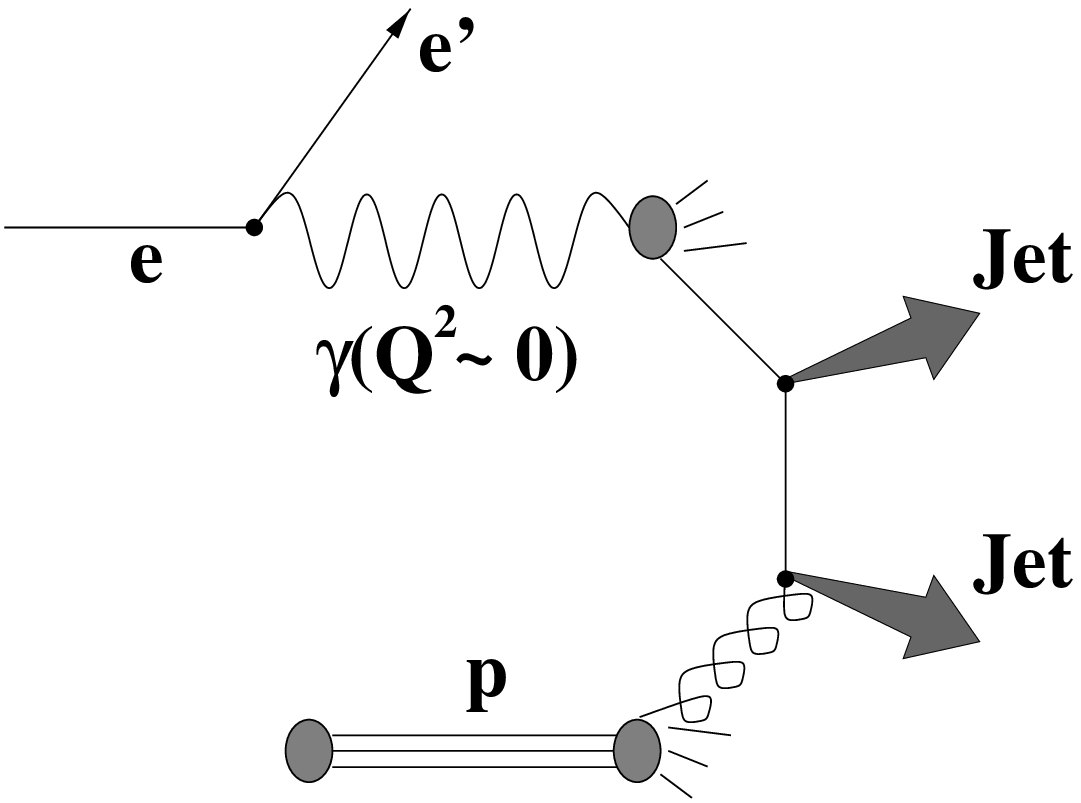,width=5cm}}
\put (4.0,0.2){\epsfig{figure=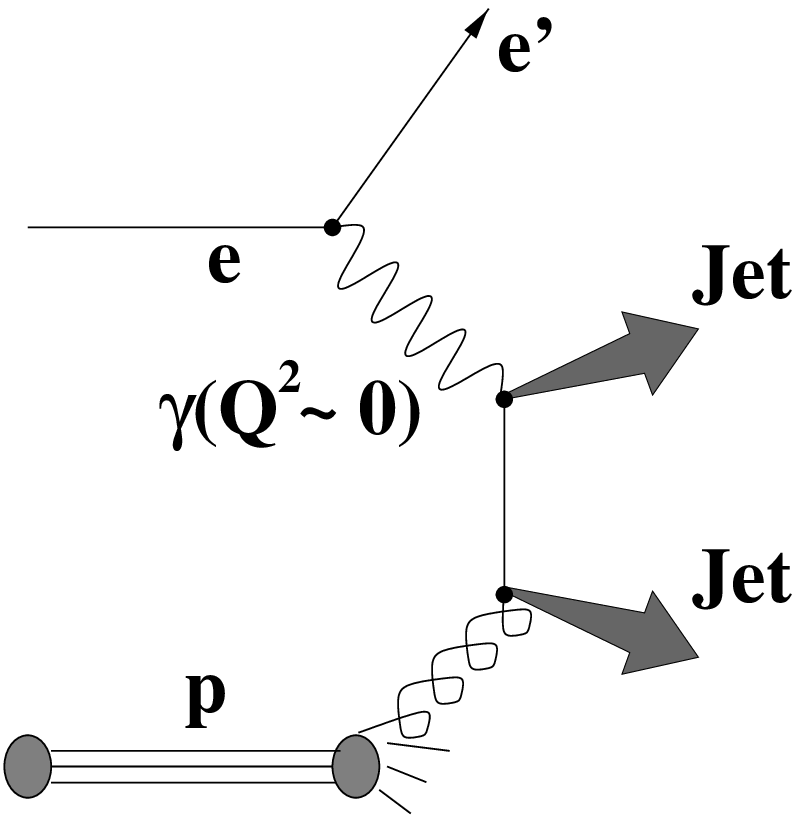,width=4cm}}
\put (-0.2,-0.3){\small (a)}
\put (5.5,-0.3){\small (b)}
\end{picture}
\end{center}
\caption{\label{fig1} (a) Resolved- and (b) direct-photon processes.}
\end{figure*}

At HERA, positrons of energy $E_e=27.5$~GeV collide with protons of energy
$E_p=820$~GeV.
The main source of jets at HERA is hard scattering in $\gp$
interactions in which a quasi-real photon ($\q2\approx 0$, where $\q2$ is the
virtuality of the photon) emitted by the positron beam
interacts with a parton from the proton to produce two jets in the final state.
At leading order (LO) QCD, there are two processes which contribute to the jet
production cross section: the resolved process (figure~\ref{fig1}a) in which
the photon interacts through its partonic content, and the direct process
(figure~\ref{fig1}b) in which the photon interacts as a point-like particle.

The cross section for jet production at LO QCD in $\gp$ interactions is
given by

\vspace{0.25cm}
\leftline{$\sigma^{LO,ep\rightarrow 2{\rm jet}}_{\rm direct}=\int d\Omega\ f_{\gamma/e}(y)\ f_{j/p}(x_p,\mu^2_F)\ \times$}
\vspace{0.25cm}
\rightline{$d\sigma(\gamma j\rightarrow {\rm jet}\ {\rm jet})$}
\vspace{0.25cm}

and

\vspace{0.25cm}
\leftline{$\sigma^{LO,ep\rightarrow 2{\rm jet}}_{\rm resolved}=\int d\Omega\ f_{\gamma/e}(y)\ f_{j/p}(x_p,\mu^2_F)\ \times$}
\vspace{0.25cm}
\rightline{$f_{i/\gamma}(x_{\gamma},\mu^2_F)
\ d\sigma(ij\rightarrow {\rm jet}\ {\rm jet}),$}
\vspace{0.25cm}

\parindent 0em
where $f_{\gamma /e}(y)$ is the flux of photons in the positron,
usually estimated by the Weizs\"acker-Williams approximation~\cite{wwa}
($y$ is the fraction of the positron energy taken by the photon);
$f_{j/p}(x_p,\mu^2_F)$ are the parton densities in the proton, determined
from e.g. global fits~\cite{mrs} ($x_p$ is the fraction of the proton
momentum taken by parton $j$ and $\mu_F$ is the factorisation scale); and
$d\sigma(\gamma(i)j\rightarrow {\rm jet}\ {\rm jet})$ is the subprocess
cross section, calculable in perturbative QCD. In the case of resolved
processes, there is an additional ingredient:
$f_{i/\gamma}(x_{\gamma},\mu^2_F)$ are the parton densities in the photon,
for which there is only partial information ($x_{\gamma}$ is the fraction
of the photon momentum taken by parton $i$). The integrals are performed over
the phase space represented by ``$d\Omega$''.
\parindent 0.8em

\subsection{The iterative cone algorithm}
In hadronic type interactions, jets are usually reconstructed by a cone
algorithm~\cite{cone}. Experimentally, jets are found in the pseudorapidity
($\eta$) $-$ azimuth ($\varphi$) plane using the transverse energy flow of the
event. The jet variables are defined according to the Snowmass
Convention~\cite{snow},

\vspace{0.25cm}
\centerline{$\etjet = \sum_i E^i_T$}
\vspace{0.25cm}
\centerline{$\etajet = \frac{\sum_i E^i_T\cdot\eta_i}{\etjet};\
\phijet = \frac{\sum_i E^i_T \cdot\varphi_i}{\etjet}$.}
\vspace{0.25cm}
In the iterative cone algorithm, jets are found by maximising the
summed transverse energy within a cone of radius $R$.

\section{Inclusive jet cross sections}
Inclusive jet cross sections have been measured~\cite{incjet} using the
$1995-1997$ ZEUS~\cite{status} data (which amounts to an integrated
luminosity of ${\cal L}\sim 43$ \pb1) as a function of the jet transverse
energy. The jets have been searched for with an iterative cone
algorithm with $\rr1$. The measurements have been performed for jets of
hadrons with $\etjet$ between 17 and 74 GeV and $\etajet$ between $-0.75$
and $2.5$, and are given for the kinematic region defined by $0.2<y<0.85$
and $\q2\leq 4$~\g2.

\begin{figure*}
\begin{center}
\setlength{\unitlength}{1.0cm}
\begin{picture} (10.0,8.0)
\put (-2.0,0.5){\epsfig{figure=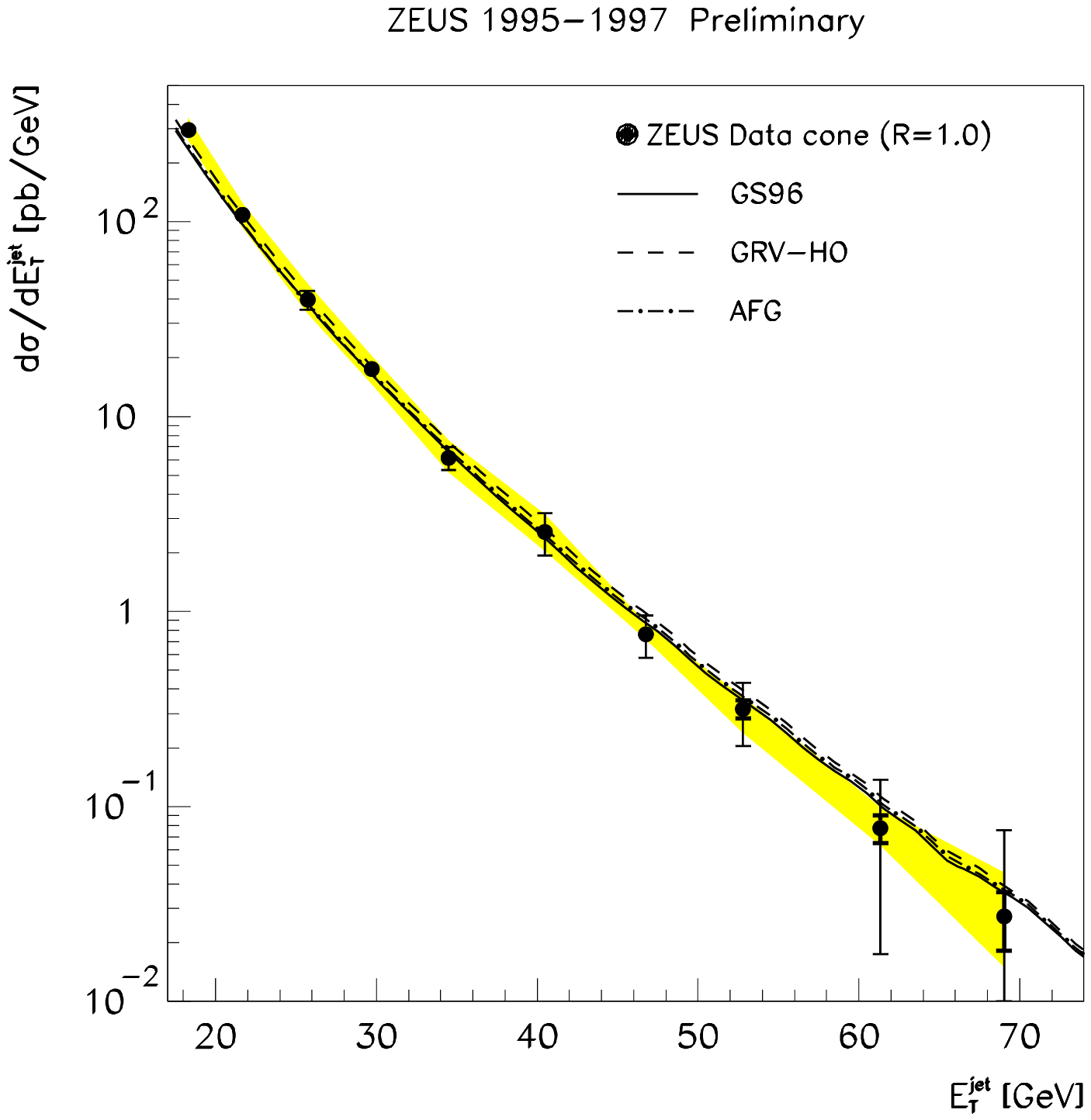,width=8cm}}
\put (4.5,0.5){\epsfig{figure=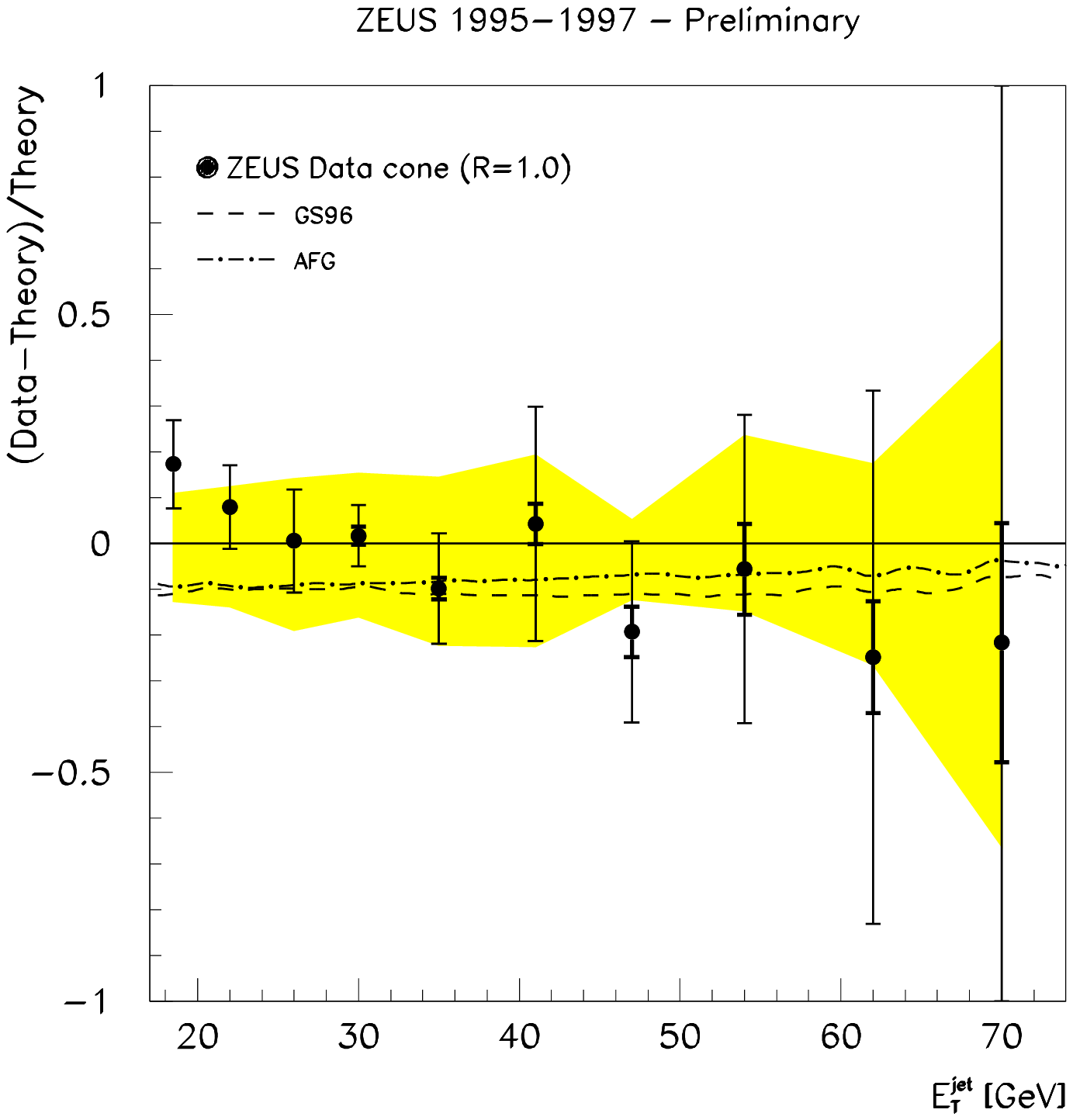,width=8cm}}
\put (3.5,6.0){\small (a)}
\put (9.5,6.0){\small (b)}
\end{picture}
\end{center}
\vspace{-1.5cm}
\caption{\label{fig2} Inclusive jet cross section as a function of $\etjet$
using the iterative cone algorithm. NLO QCD calculations are shown for
comparison.}
\end{figure*}

Figure~\ref{fig2}a shows the measured $\set$ (black dots). The systematic
uncertainties not associated with the absolute energy scale of the jets have
been added in quadrature to the statistical errors (thick error bars) and are
shown as thin error bars. The shaded band represents the uncertainty on the
energy scale of the jets. The data show a steep fall-off over four orders of
magnitude in the measured range.

\subsection{NLO QCD calculations}
There are several complete calculations of jet cross sections at NLO for $\gp$
interactions~\cite{klasen1,harris,frixione,aurenche}. Two types of
corrections contribute at NLO: the virtual corrections which include internal
particle loops and the real corrections which include a third parton in the
final state. The existing calculations differ mainly in the treatment of the
real corrections. A detailed comparison of the calculations~\cite{khf} show
that for NLO calculations of dijet cross sections and LO calculations of
three-jet cross sections a reasonable agreement is found, and the differences
are up to $\sim 5\%$.

The curves in figure~\ref{fig2}a are NLO QCD
calculations~\cite{klasen1,harris} using different parametrisations of the
photon structure function: GS96~\cite{gs} (solid line), GRV-HO~\cite{grv}
(dashed line) and AFG~\cite{afg} (dot-dashed line). The CTEQ4M~\cite{cteq4}
proton parton densities have been used in all cases. In the calculations shown
here, the renormalisation and factorisation scales have been chosen equal to
$\etjet$ and $\alpha_s$ was calculated at 2-loops with
$\Lambda^{(4)}_{\overline{MS}}=296$ MeV. The NLO calculations give a
reasonable description of the data. Figure~\ref{fig2}b shows the fractional
differences between the measured $\set$ and the NLO calculations based on
GRV-HO.

\subsection{Comparing theory and experiment}
To perform tests of QCD and to extract information on the photon parton
densities, the experimental and theoretical uncertainties must be
reduced as much as possible.

Among the main experimental uncertainties the presence of a
possible underlying event, which is the result of soft interactions between the
partons in the photon and proton remnants (for resolved events), and is not
included in the calculations. The uncertainty of the measurements due to the
underlying event is reduced by decreasing the cone radius or by increasing the
transverse energy of the jets~\cite{zeus2}.
%

On the theoretical side, since calculations are made only at NLO, the
implementation of the iterative cone jet algorithm in the theory does not
match the experimental procedure exactly. The theoretical uncertainty coming
from this effect is reduced by using the longitudinally invariant $\kt$
cluster algorithm~\cite{kt}.

\subsection{The $\kt$ cluster algorithm}
In the inclusive $\kt$ cluster algorithm~\cite{kt}, jets are identified by
successively combining nearby pairs of particles until a jet is complete. The
$\kt$ algorithm allows a transparent translation of the theoretical jet
algorithm to the experimental set-up by avoiding the ambiguities related to
the merging and overlapping of jets and it is infrared safe to all orders.

\begin{figure*}
\begin{center}
\setlength{\unitlength}{1.0cm}
\begin{picture} (10.0,8.0)
\put (-2.0,0.5){\epsfig{figure=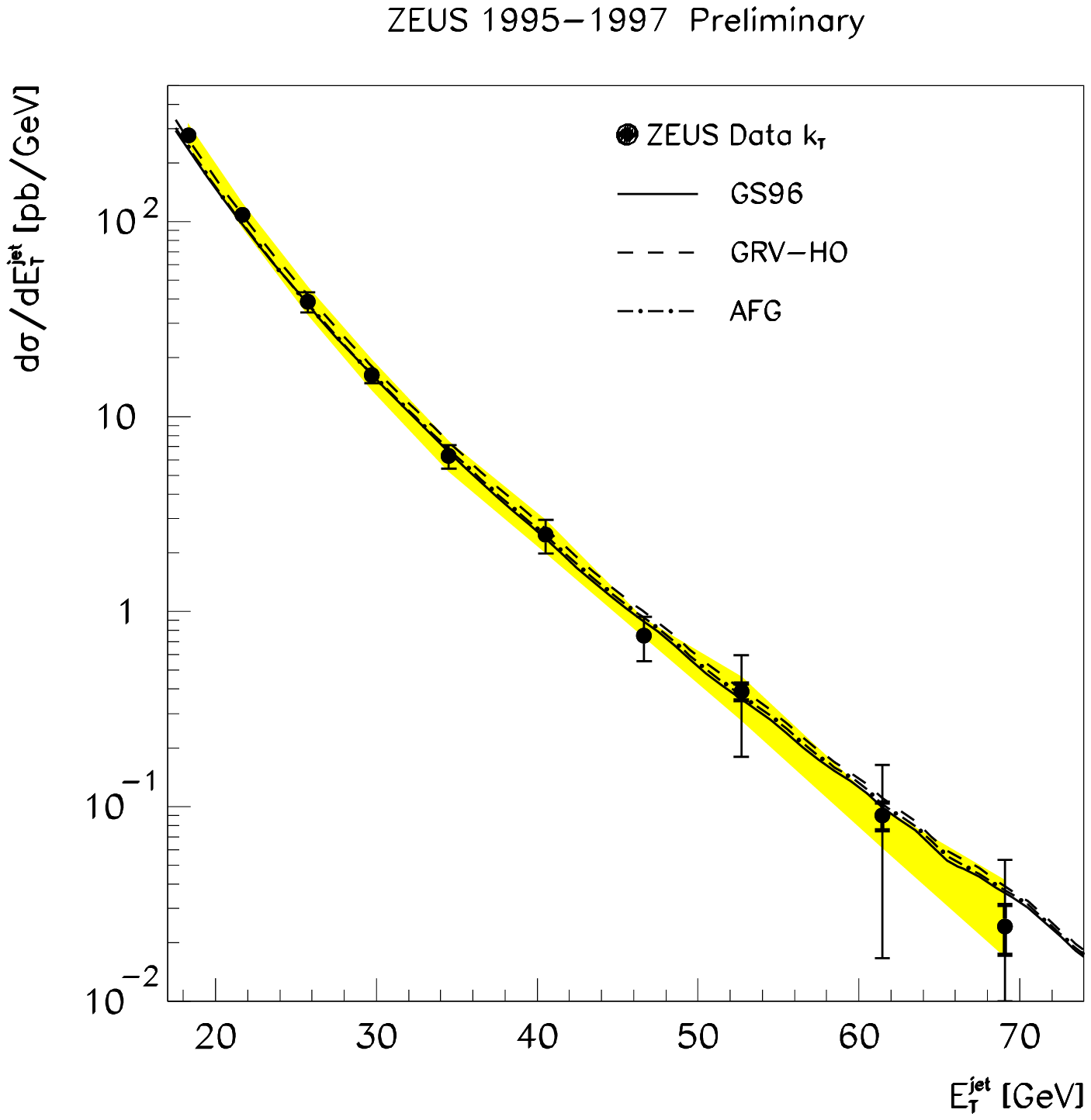,width=8cm}}
\put (4.5,0.5){\epsfig{figure=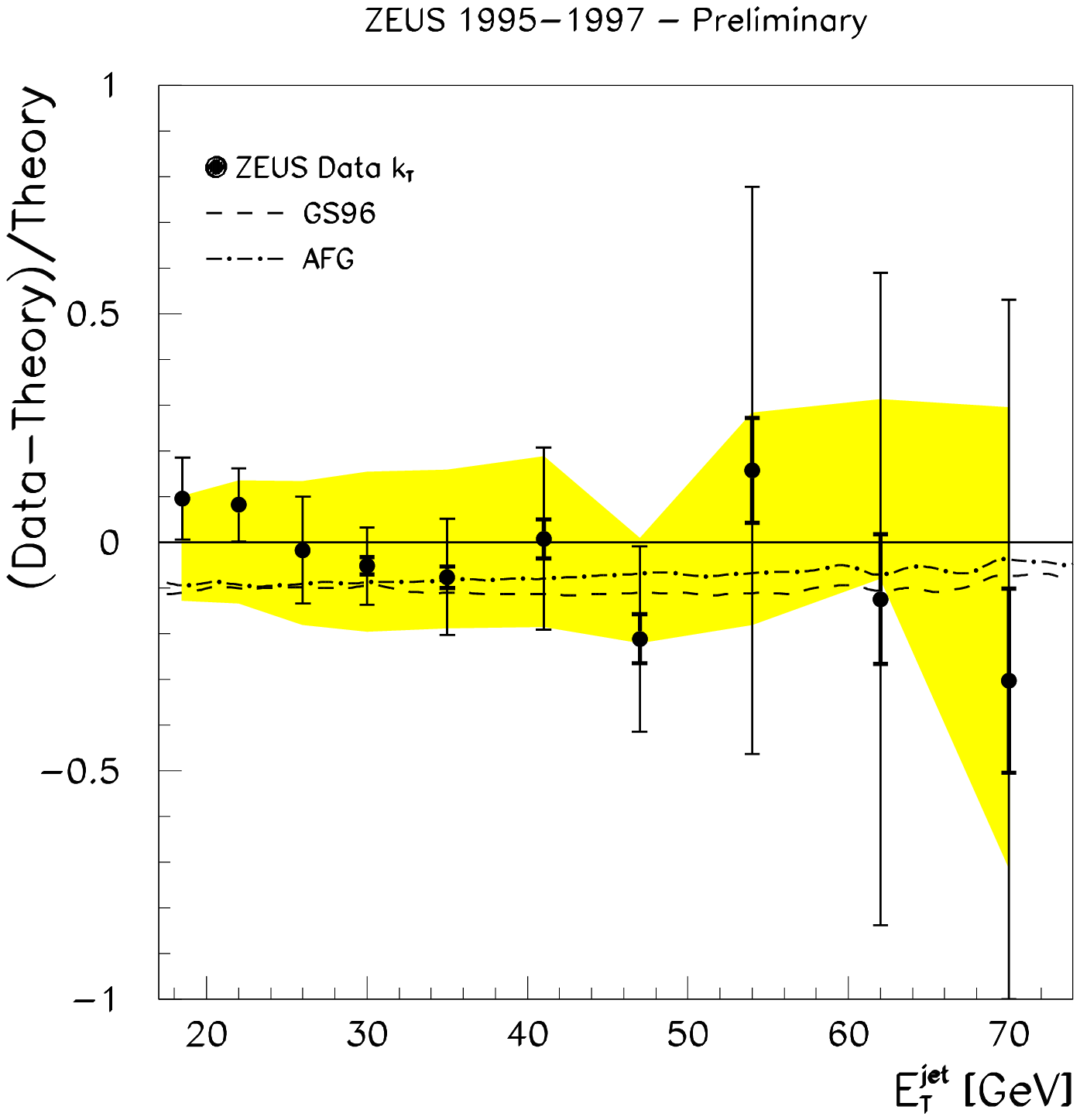,width=8cm}}
\put (3.5,6.5){\small (a)}
\put (10.0,6.5){\small (b)}
\end{picture}
\end{center}
\vspace{-1.5cm}
\caption{\label{fig3}  Inclusive jet cross section as a function of $\etjet$
using the $\kt$ cluster algorithm. NLO QCD calculations are shown for
comparison.}
\end{figure*}

Figure~\ref{fig3} shows $\set$ for jets found using the $\kt$ cluster
algorithm. The NLO QCD calculations, using the current knowledge of the photon
structure, are able to describe the data within the present experimental and
theoretical uncertainties.

\begin{figure*}
\begin{center}
\setlength{\unitlength}{1.0cm}
\begin{picture} (10.0,8.0)
\put (1.5,0.5){\epsfig{figure=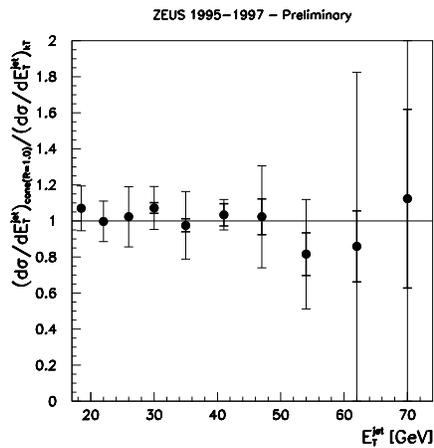,width=8cm}}
\end{picture}
\end{center}
\vspace{-1.5cm}
\caption{\label{fig4} Ratio between the measured $\set$ using the iterative
cone and the $\kt$ cluster algorithms.}
\end{figure*}

A comparison between $\set$ measured using the cone and the $\kt$ algorithms
has been made (see figure~\ref{fig4}). The differences between the measured
cross sections are typically smaller than 10\%. This comparison shows
that the cone with $\rr1$ and the $\kt$ cluster algorithms probe the
underlying parton dynamics in a comparable way. Therefore, in the experiment,
the choice of jet algorithm is not crucial. The use of the $\kt$ cluster
algorithm is dictated by the need to reduce the theoretical uncertainties.

\section{Jet substructure}
The use of the $\kt$ cluster algorithm allows the study of the internal
structure of the jets in terms of subjets. Subjets are jet-like objects within
a jet and are resolved by reapplying the $\kt$ cluster algorithm until for
every pair of particles $i$ and $j$,

\vspace{0.25cm}
\leftline{$d_{ij}=\min(E_{Ti},E_{Tj})^2[(\eta_i-\eta_j)^2+(\varphi_i-\varphi_j)^2]$}
\vspace{0.25cm}
\rightline{$\geq y_{\rm cut}(\etjet)^2$.}
\vspace{0.25cm}

Measurements of the mean subjet multiplicity ($<n_{\rm subjet}>$) have been
performed by ZEUS~\cite{subjet} using an inclusive sample of jets with
$\etjet>15$~GeV and $-1<\etajet<2$ in the kinematic range defined by
$0.2<y<0.85$ and $\q2\leq 1$~\g2.

\begin{figure*}
\begin{center}
\setlength{\unitlength}{1.0cm}
\begin{picture} (10.0,8.0)
\put (-2.0,0.5){\epsfig{figure=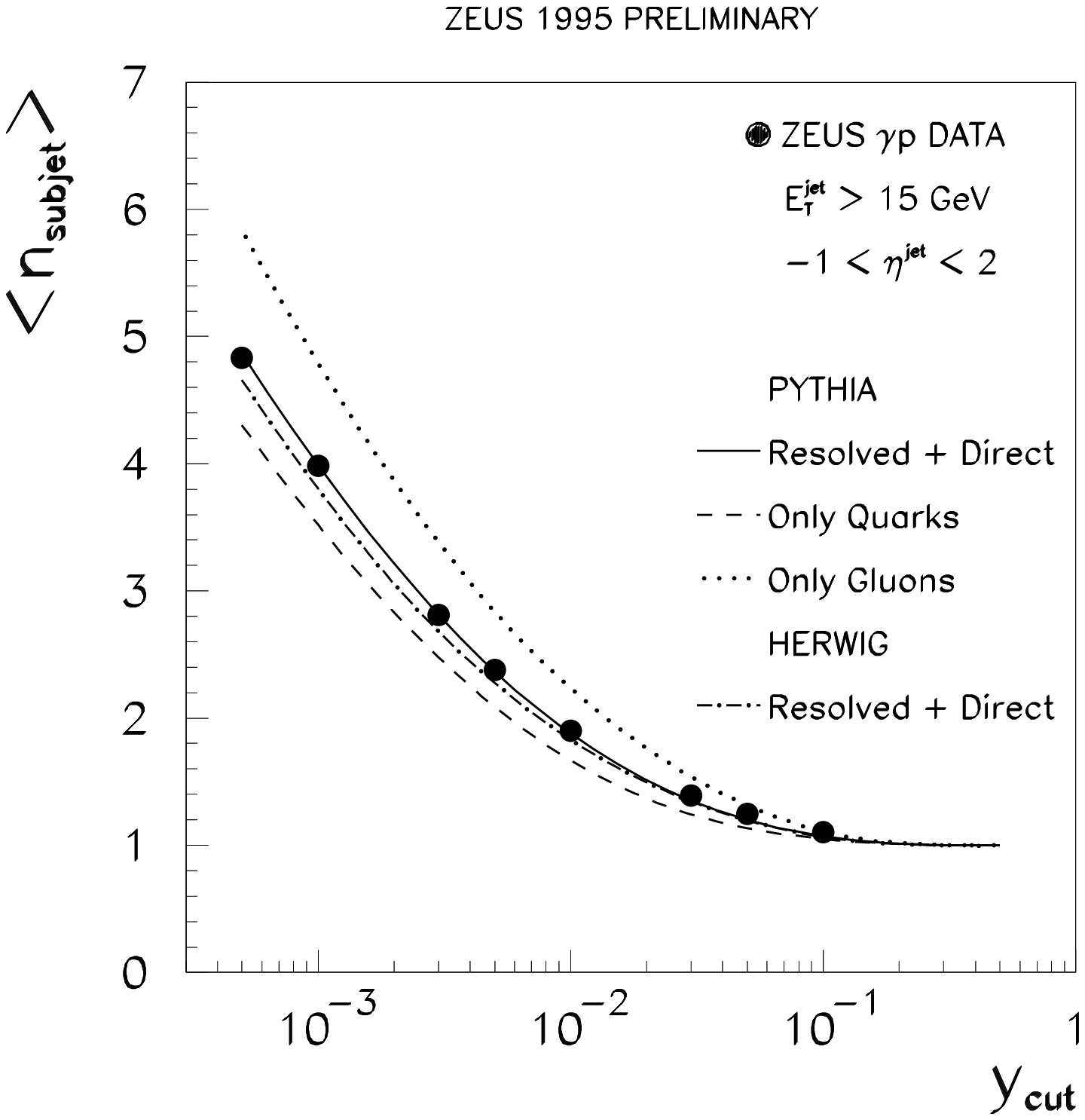,width=8cm}}
\put (4.5,0.5){\epsfig{figure=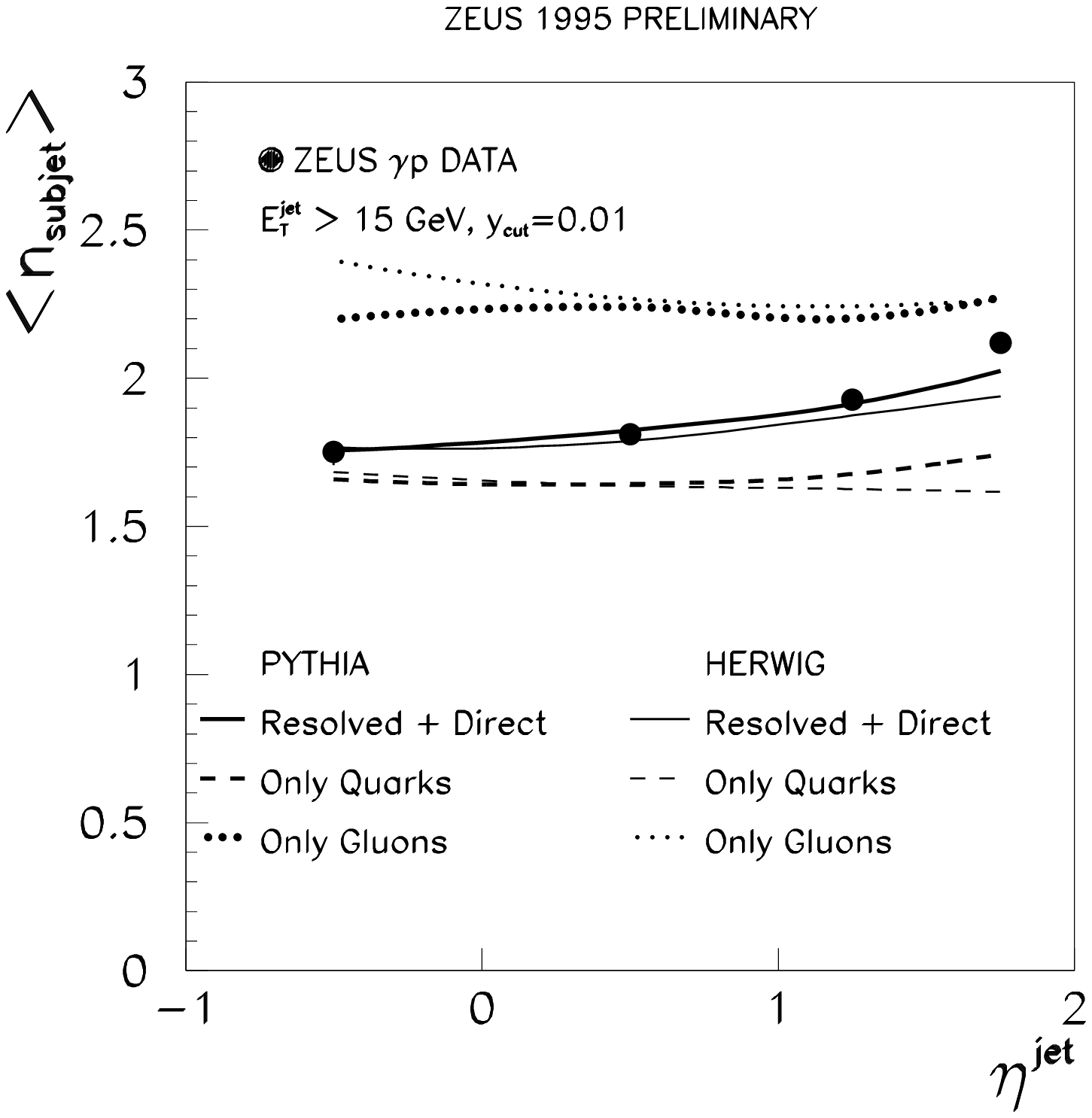,width=8cm}}
\put (0.5,6.5){\small (a)}
\put (10.0,6.5){\small (b)}
\end{picture}
\end{center}
\vspace{-1.5cm}
\caption{\label{fig5} Mean subjet multiplicity as a function of (a)
$y_{\rm cut}$ for $-1<\etajet<2$ and (b) $\etajet$ for $y_{\rm cut}=0.01$.
PYTHIA and HERWIG Monte Carlo calculations are shown for comparison.}
\end{figure*}

Figure~\ref{fig5}a shows $<n_{\rm subjet}>$ as a function of
$y_{\rm cut}$. The data (black dots, the statistical and systematic
uncertainties are included but are smaller than the dots) show that
$<n_{\rm subjet}>$ grows as $y_{\rm cut}$ decreases within the measured
range. The lines are calculations from the leading-logarithm parton-shower
Monte Carlo's PYTHIA~\cite{pythia} and HERWIG~\cite{herwig}. The
calculations based on PYTHIA give a good description of the data.
Figure~\ref{fig5}b shows $<n_{\rm subjet}>$ as a function of $\etajet$ for
$y_{\rm cut}=0.01$: $<n_{\rm subjet}>$ increases as $\etajet$ increases.
The comparison with the predictions for quark and gluon jets shows that the
increase in $<n_{\rm subjet}>$ as $\etajet$ increases is consistent with
the predicted increase in the fraction of gluon jets.

\section{Dijet cross sections}
Dijet cross sections have been measured~\cite{dijeth1} using the
$1994-1997$ H1~\cite{statush1} data (which amounts to an integrated
luminosity of ${\cal L}\sim 36$ \pb1) as a function of the transverse
energy of the leading jet and the average transverse
energy of the two leading jets. The jets have
been found using the $\kt$ cluster algorithm. The measurements have been
performed for jets of hadrons with $E_T^{jet1}>25$~GeV, $E_T^{jet2}>15$~GeV
and $-0.5<\etajet<2.5$, and are given for the kinematic region defined by
$y<0.9$ and $\q2<4$~\g2.

\begin{figure*}
\begin{center}
\setlength{\unitlength}{1.0cm}
\begin{picture} (10.0,8.0)
\put (-2.0,1.5){\epsfig{figure=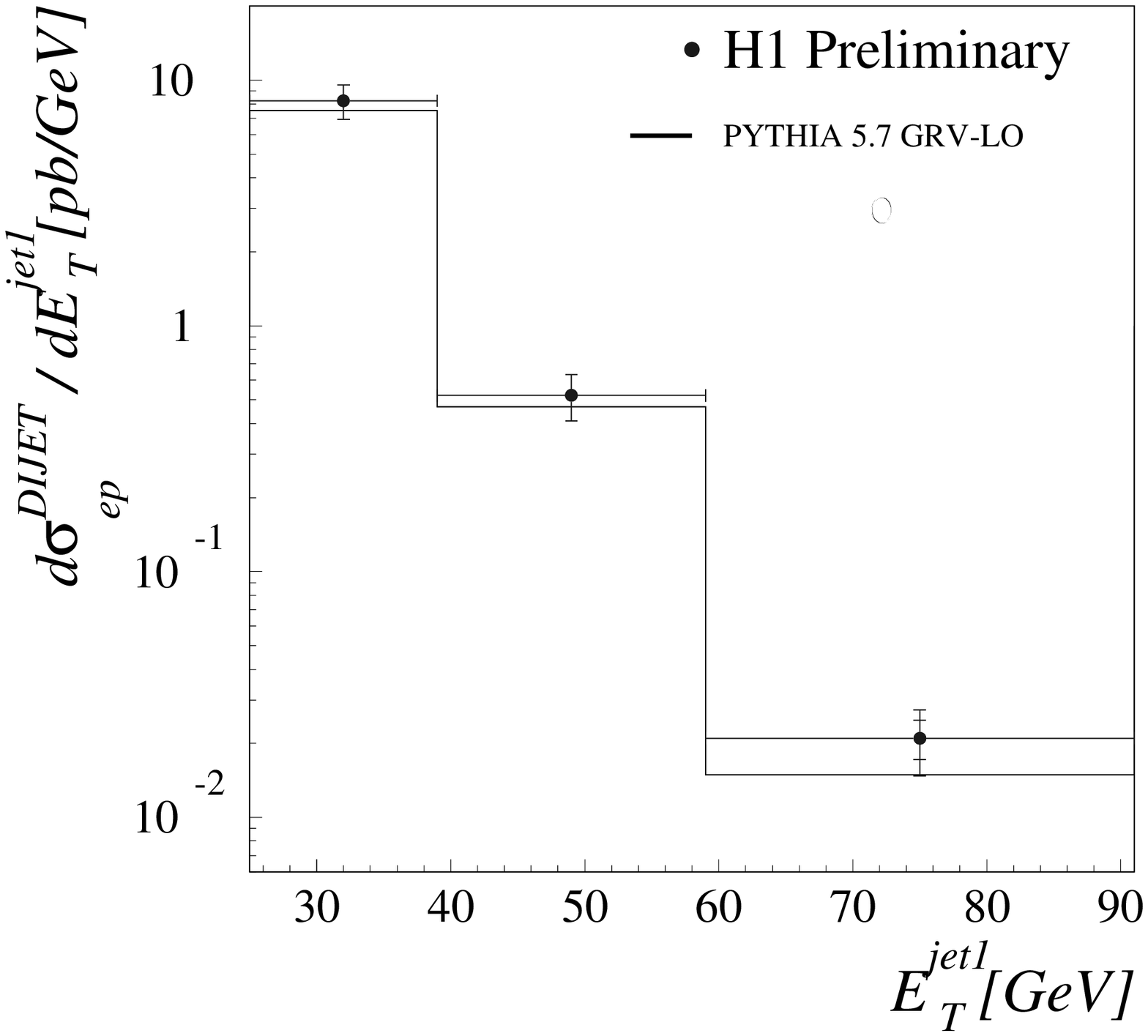,width=6.7cm}}
\put (4.5,1.5){\epsfig{figure=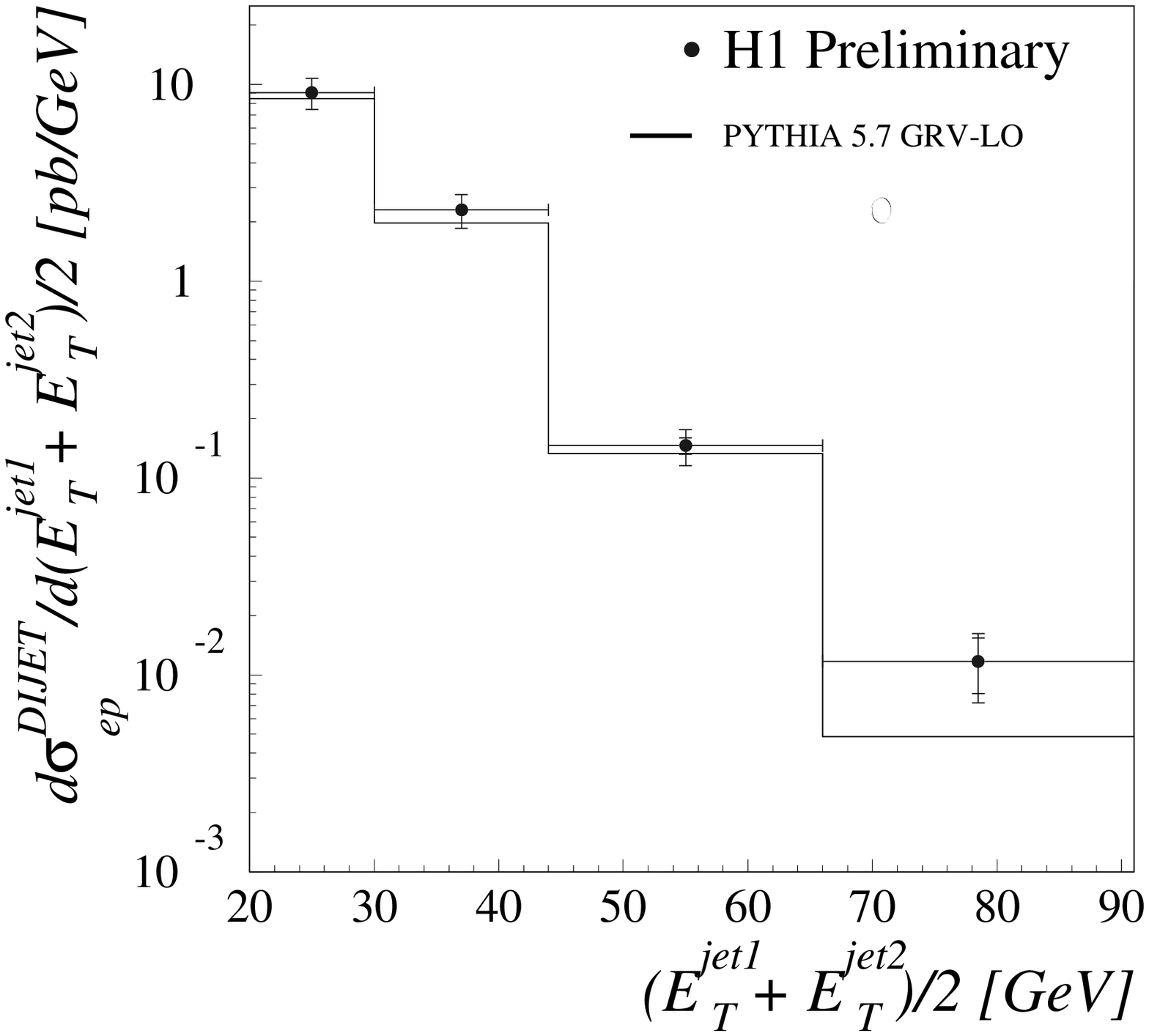,width=6.7cm}}
\put (3.5,6.0){\small (a)}
\put (9.5,6.0){\small (b)}
\end{picture}
\end{center}
\vspace{-2cm}
\caption{\label{fig6} Dijet cross sections as a function of (a) the
transverse energy of the leading jet and (b) the average transverse energy
of the two leading jets. PYTHIA Monte Carlo calculations are shown for
comparison.}
\end{figure*}

Figure~\ref{fig6} shows the measured cross sections (black dots). The
systematic uncertainties (including that associated with the absolute energy
scale of the jets) have been added in quadrature to the statistical errors
(inner error bars) and are shown as the outer error bars. The data show a
steep fall-off of three orders of magnitude in the measured range. The
histograms are the calculations using PYTHIA which provide a
good description of the shape of the measured distributions.

\section{High-mass dijet cross sections}
The dijet mass distribution $\mj$ is sensitive to the presence of new
particles or resonances that decay into two jets. The distribution of the
angle between the jet-jet axis and the beam direction in the dijet
centre-of-mass system ($\cos\theta^*$) reflects the underlying parton dynamics
and is sensitive to the spin of the exchanged particle. New particles or
resonances decaying into two jets may also be identified by deviations in the
measured $\cos\theta^*$ distribution with respect to the QCD predictions.

\begin{figure*}
\begin{center}
\setlength{\unitlength}{1.0cm}
\begin{picture} (10.0,8.0)
\put (2.0,1.5){\epsfig{figure=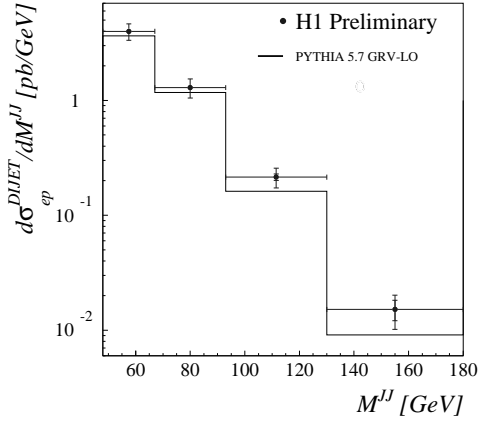,width=6.7cm}}
\end{picture}
\end{center}
\vspace{-2cm}
\caption{\label{fig7} Dijet cross section as a function of the dijet
invariant mass. PYTHIA Monte Carlo calculations are shown for comparison.}
\end{figure*}

The cross section as a function of the dijet invariant mass has been
measured by H1~\cite{dijeth1} using the $\kt$ cluster algorithm between 48
and 148 GeV (figure~\ref{fig7}). The data show a steep fall off of more than
two orders of magnitude within the measured range. The histogram is the
calculation using PYTHIA which gives a reasonable description of the shape of
the measured distribution. There are large uncertainties in the normalisation
of the LO QCD calculations which indicate the need for NLO corrections.

High-mass dijet cross sections have been measured~\cite{dijetze} using the
$1995-1997$ ZEUS data as a function of $\mj$ and $\cost$. The measurements
have been performed for $\mj>47$ GeV and $\cost<0.8$ using the $\kt$ cluster
algorithm.

\begin{figure*}
\begin{center}
\setlength{\unitlength}{1.0cm}
\begin{picture} (10.0,8.0)
\put (-2.0,0.5){\epsfig{figure=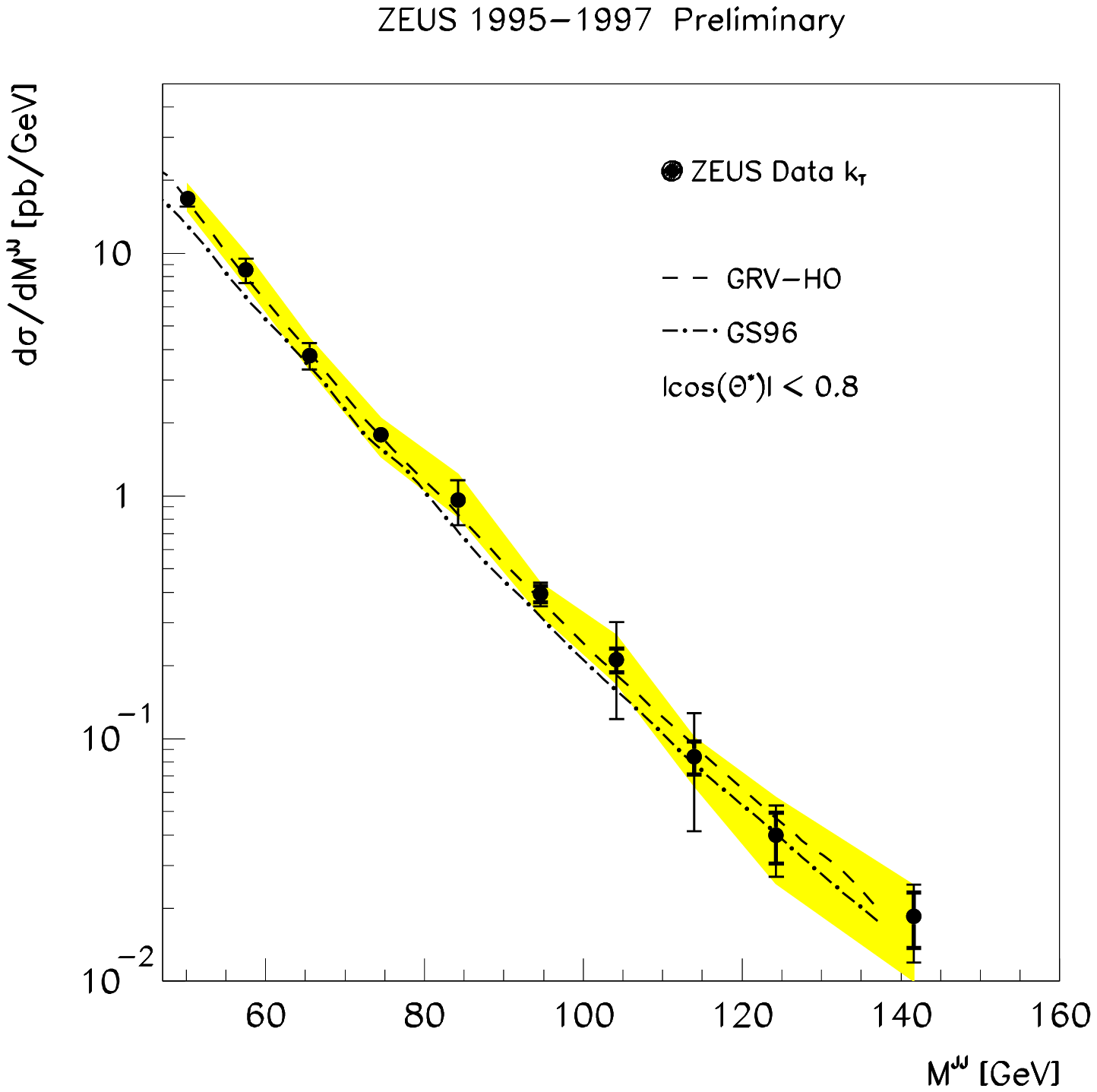,width=8cm}}
\put (4.5,0.5){\epsfig{figure=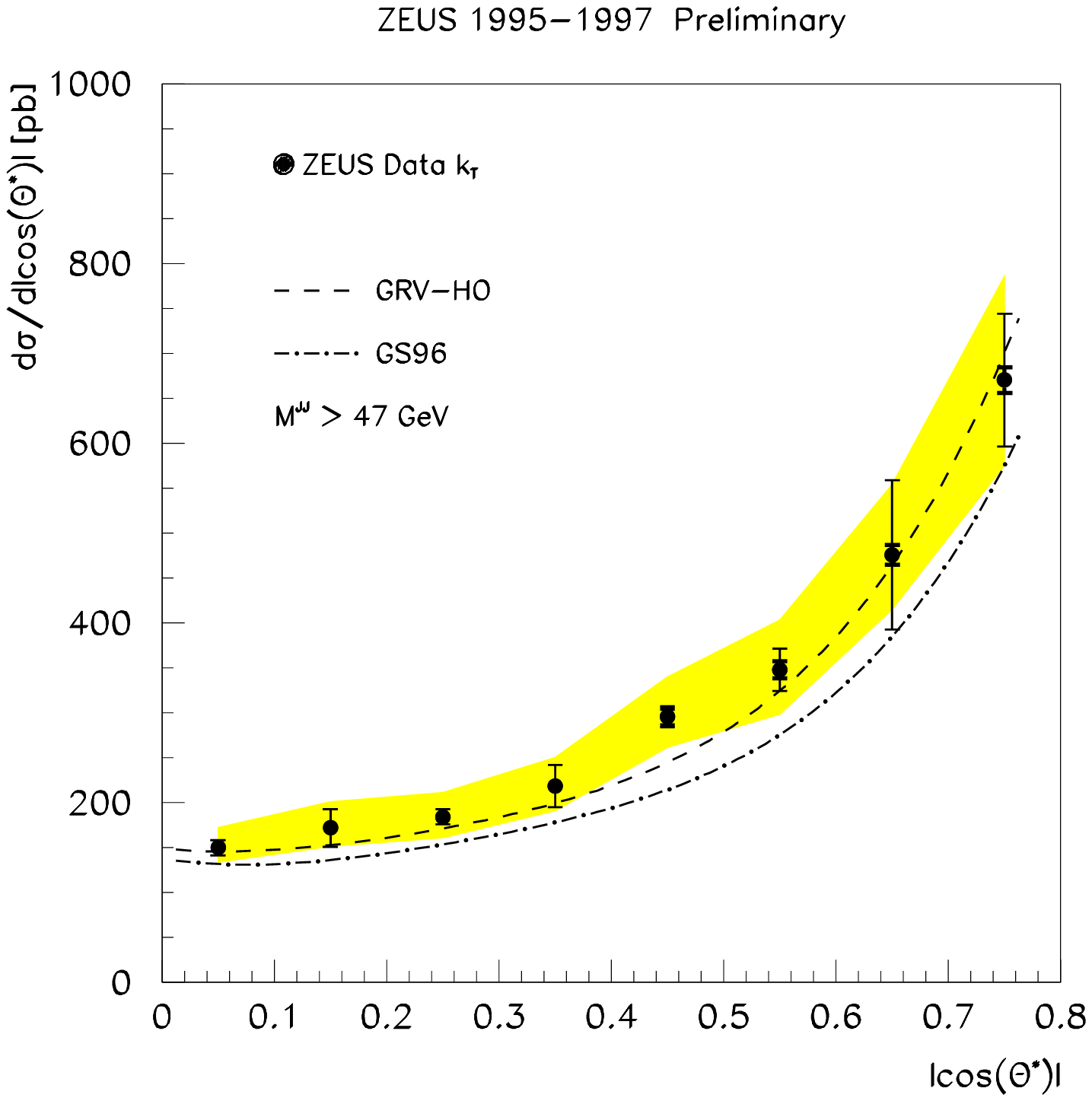,width=8cm}}
\put (3.5,6.0){\small (a)}
\put (9.5,6.0){\small (b)}
\end{picture}
\end{center}
\vspace{-1.5cm}
\caption{\label{fig8} High-mass dijet cross sections as a function of (a)
the dijet invariant mass and (b) the dijet angular distribution.
NLO QCD calculations are shown for comparison.}
\end{figure*}

The data show a steep fall-off in $\mj$ of three orders of magnitude in the
measured range (figure~\ref{fig8}a). The measured $\scost$ rises as $\cost$
increases (figure~\ref{fig8}b). The NLO QCD calculations~\cite{klasen1} give
a reasonable description of the measured distributions. The calculations based
on GRV-HO are closer in magnitude to the measured cross sections. No
significant deviation between data and NLO calculations is observed in the
measured range of $\mj$ and $\cost$.

\begin{figure*}
\begin{center}
\setlength{\unitlength}{1.0cm}
\begin{picture} (10.0,6.0)
\put (2.3,1.0){\epsfig{figure=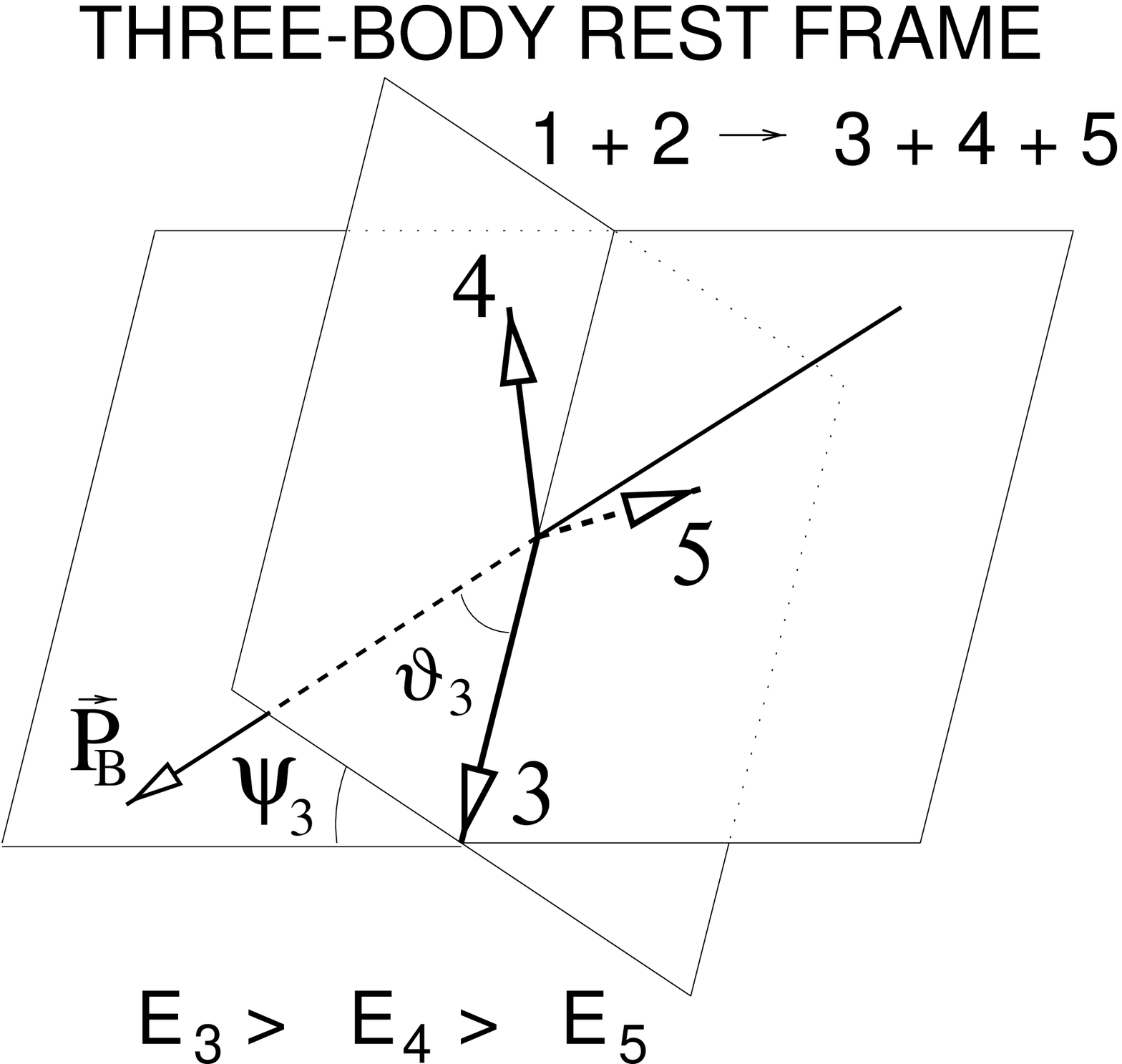,width=6.4cm}}
\put (2.5,6.63){\epsfig{figure=white.eps,width=6cm,height=0.5cm}}
\end{picture}
\end{center}
\vspace{-1.5cm}
\caption{\label{fig9} Three-body rest frame.}
\end{figure*}

\section{Three-jet cross sections}
Measurements of three-jet cross sections provide a test of QCD beyond
LO and allow a search for new phenomena. NLO calculations for three-jet
cross sections in $\gp$ interactions are not yet available. The calculations
shown here are LO for these processes and are therefore subject to large
renormalisation and factorisation scale uncertainties. The cross section for
three-jet production at LO is given by

\vspace{0.25cm}
\leftline{$\sigma^{LO,ep\rightarrow 3{\rm jet}}_{\rm dir\ [res]}=\int d\Omega\ \
f_{\gamma /e}(y)\ \ f_{j/p}(x_p,\mu^2_F)\ \times$}
\vspace{0.25cm}
\rightline{$[f_{i/\gamma}(x_{\gamma},\mu^2_F)]\ \
d\sigma(\gamma[i]j\rightarrow {\rm jet}\ {\rm jet}\ {\rm jet}).$}
\vspace{0.25cm}

Five parameters are necessary to uniquely determine the three-body phase-space.
These are the three-jet invariant mass ($M_{3j}$); the energy-sharing
quantities $X_3$ and $X_4$ (the jets are numbered 3, 4 and 5 in order of
decreasing energy),

\vspace{0.25cm}
\centerline{$X_i\equiv\frac{2E_i}{M_{3j}}$;}
\vspace{0.25cm}

\parindent 0em
the cosine of the scattering angle of the highest energy jet with
respect to the beam,

\vspace{0.25cm}
\centerline{$\costh3\equiv\frac{\vec{p}_{B}\cdot\vec{p}_3}{|\vec{p}_{B}| |\vec{p}_3|}$;}
\vspace{0.25cm}

and $\psi_3$, the angle between the plane containing the highest energy jet 
and the beam and the plane containing the three jets.  The latter is defined by

\vspace{0.25cm}
\centerline{$\cos{\psi_3}\equiv\frac{(\vec{p}_3\times\vec{p}_{B})\cdot(\vec{p}_4 \times \vec{p}_5)}{|\vec{p}_3 \times \vec{p}_{B}| |\vec{p}_4 \times \vec{p}_5|}$.}
\vspace{0.25cm}

\parindent 0.8em
The definition of the angles $\theta_3$ and $\psi_3$ is illustrated in
figure~\ref{fig9}.

Since $\theta_3$ involves only the highest energy jet, the distribution of
$\costh3$ in three-jet events is expected to follow closely the
distribution of $\cos\theta^*$ in dijet events.  The $\psi_3$ angle, on the
other hand, reflects the orientation of the lowest energy jet.

\begin{figure*}
\begin{center}
\setlength{\unitlength}{1.0cm}
\begin{picture} (10.0,8.0)
\put (-2.0,1.5){\epsfig{figure=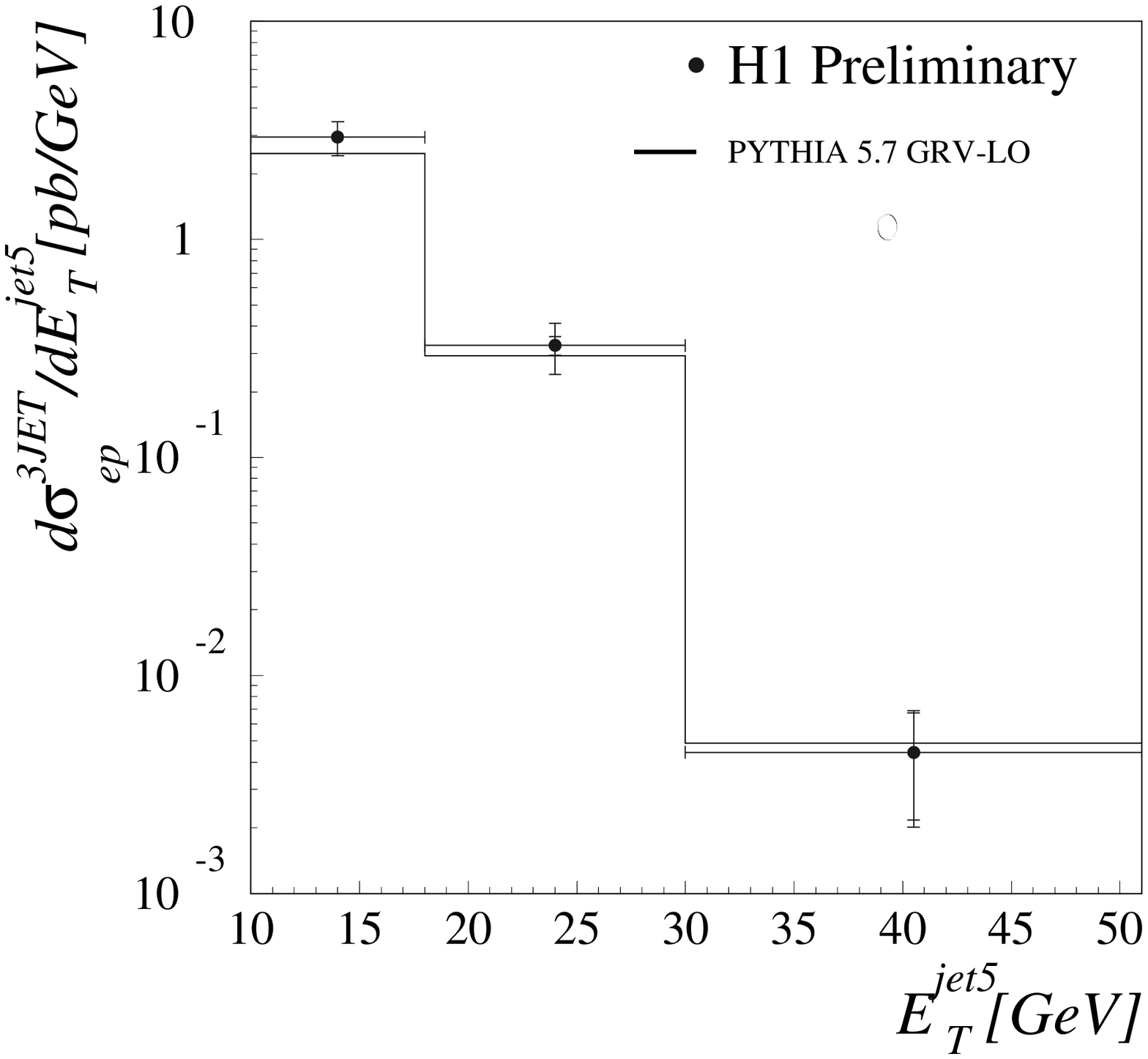,width=6.7cm}}
\put (4.5,1.5){\epsfig{figure=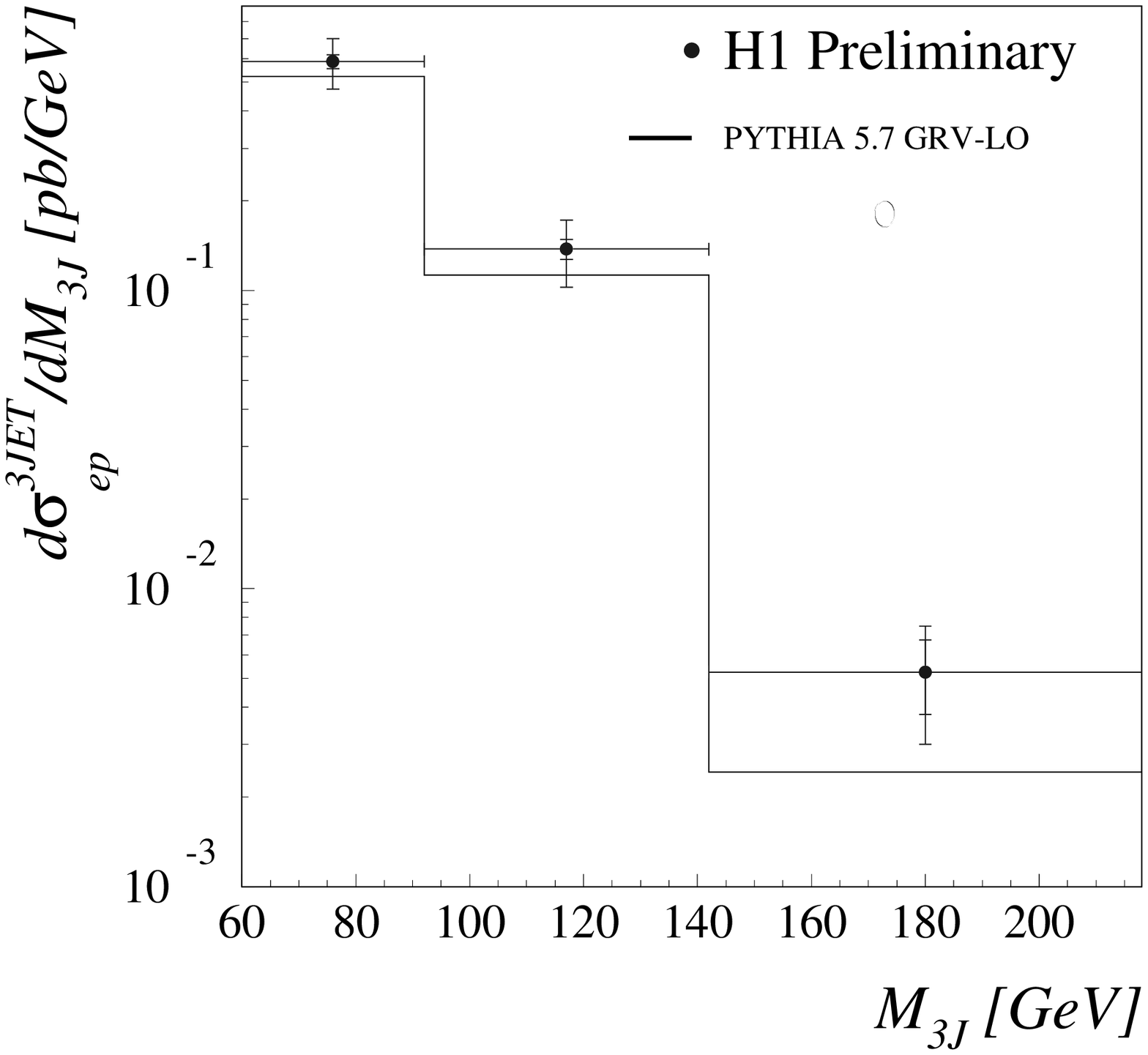,width=6.7cm}}
\put (3.0,6.0){\small (a)}
\put (9.0,6.0){\small (b)}
\end{picture}
\end{center}
\vspace{-2cm}
\caption{\label{fig10} Three-jet cross section as a function of (a) the
transverse energy of the lowest-energy jet and (b) the three-jet invariant
mass. PYTHIA Monte Carlo calculations are shown for comparison.}
\end{figure*}

Figure~\ref{fig10} shows the three-jet cross section as a function of the
transverse energy of the lowest energy jet and the three-jet invariant mass.
The comparison to the calculations using PYTHIA Monte Carlo shows that the
parton shower models describe the shape of the measured cross sections.

\begin{figure*}
\begin{center}
\setlength{\unitlength}{1.0cm}
\begin{picture} (10.0,8.0)
\put (1.7,1.0){\epsfig{figure=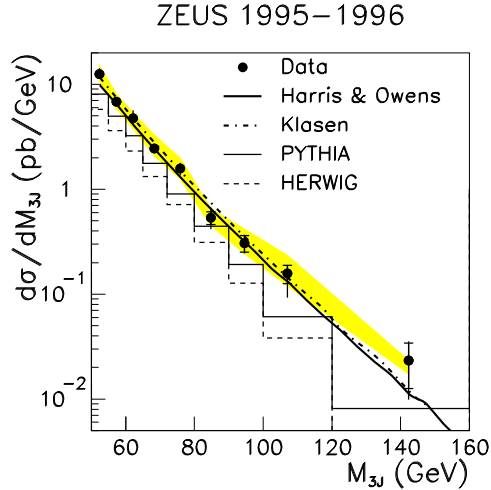,width=6.4cm}}
\end{picture}
\end{center}
\vspace{-1.5cm}
\caption{\label{fig11} Three-jet invariant mass cross section.
$\oaa$ and Monte Carlo calculations are shown for comparison.}
\end{figure*}

The three-jet invariant mass cross section was measured by
ZEUS~\cite{zeus3j} in the kinematic region defined by $|\costh3|<0.8$ and
$X_3<0.95$ (see figure~\ref{fig11}). The curves in figure~\ref{fig11} are
the $\oaa$ QCD calculations~\cite{klasenk,harrisk}
using the GRV-LO~\cite{grv} parametrisations of the photon structure function
and the CTEQ4 LO~\cite{cteq4} proton parton densities.
The renormalisation and factorisation scales have been chosen equal to
$E_T^{\max}$, the largest of the $\etjet$ values of the three jets.
$\alpha_s$ was calculated at 1-loop with
$\Lambda^{(5)}_{\overline{MS}}=181$ MeV. The calculations give a
good description of the data, even though they are LO for this
process. Monte Carlo calculations are also compared to the data: they provide
a good description of the data in shape, but the magnitude is $30-40\%$ too
low.

\begin{figure*}
\begin{center}
\setlength{\unitlength}{1.0cm}
\begin{picture} (10.0,12.8)
\put (-2.0,6.5){\epsfig{figure=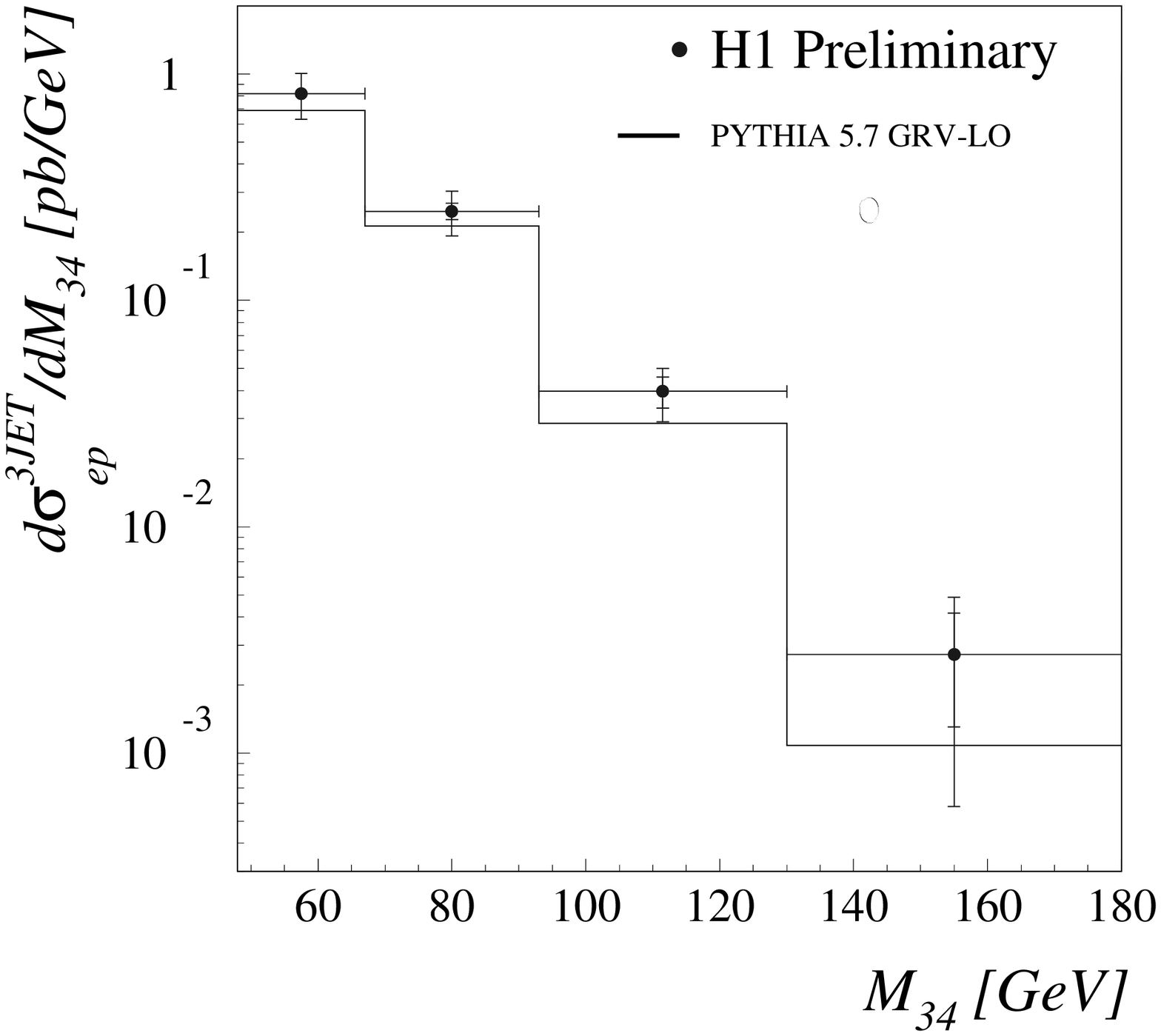,width=6.7cm}}
\put (4.5,6.5){\epsfig{figure=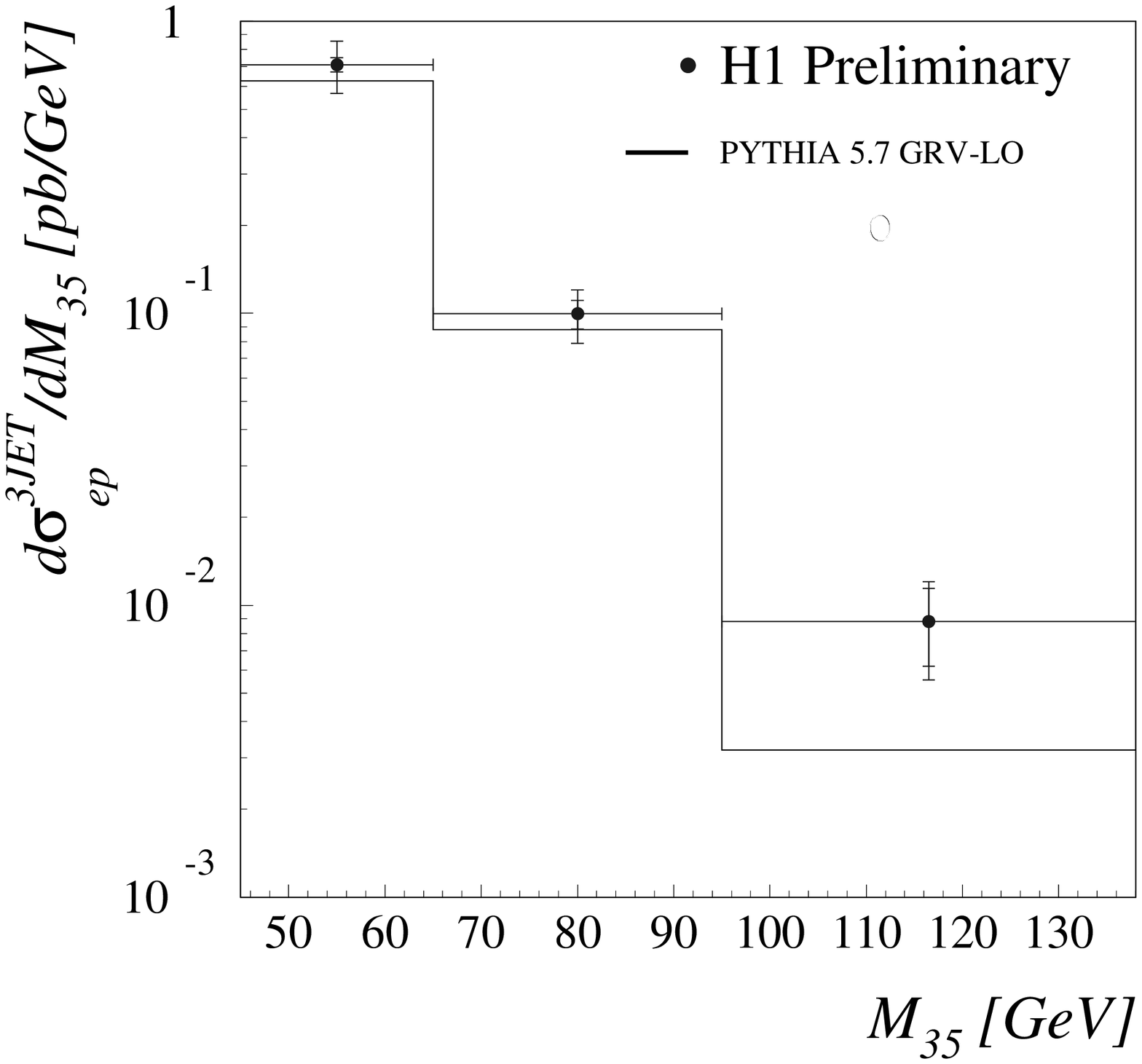,width=6.7cm}}
\put (2.0,0.5){\epsfig{figure=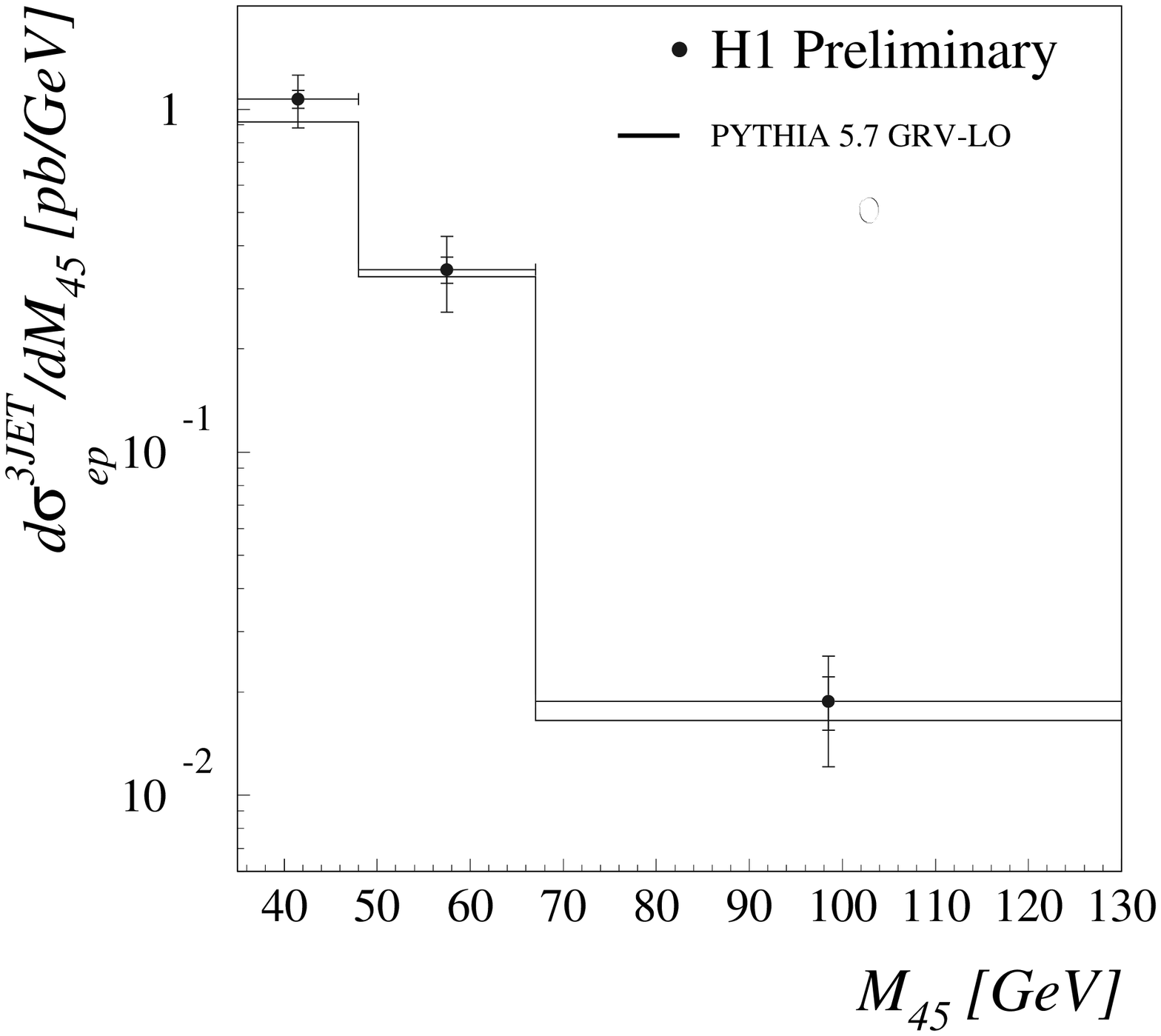,width=6.7cm}}
\put (3.0,11.0){\small (a)}
\put (9.0,11.0){\small (b)}
\put (7.0,5.0){\small (c)}
\end{picture}
\end{center}
\vspace{-1cm}
\caption{\label{fig12} Dijet invariant mass cross sections in three-jet
events. PYTHIA Monte Carlo calculations are shown for comparison.}
\end{figure*}

The search for new particles or resonances decaying into two jets can be
extended by looking for deviations in the distributions of the dijet invariant
masses in three-jet events with respect to the predictions of QCD.
Figure~\ref{fig12} shows the dijet invariant mass distributions in three-jet
events for all possible pairs of jets. The histograms are the predictions
from the QCD-based Monte Carlo models. No significant deviation between data
and calculations is observed up to the highest invariant mass value studied.

\begin{figure*}
\begin{center}
\setlength{\unitlength}{1.0cm}
\begin{picture} (10.0,12.8)
\put (0.0,0.0){\epsfig{figure=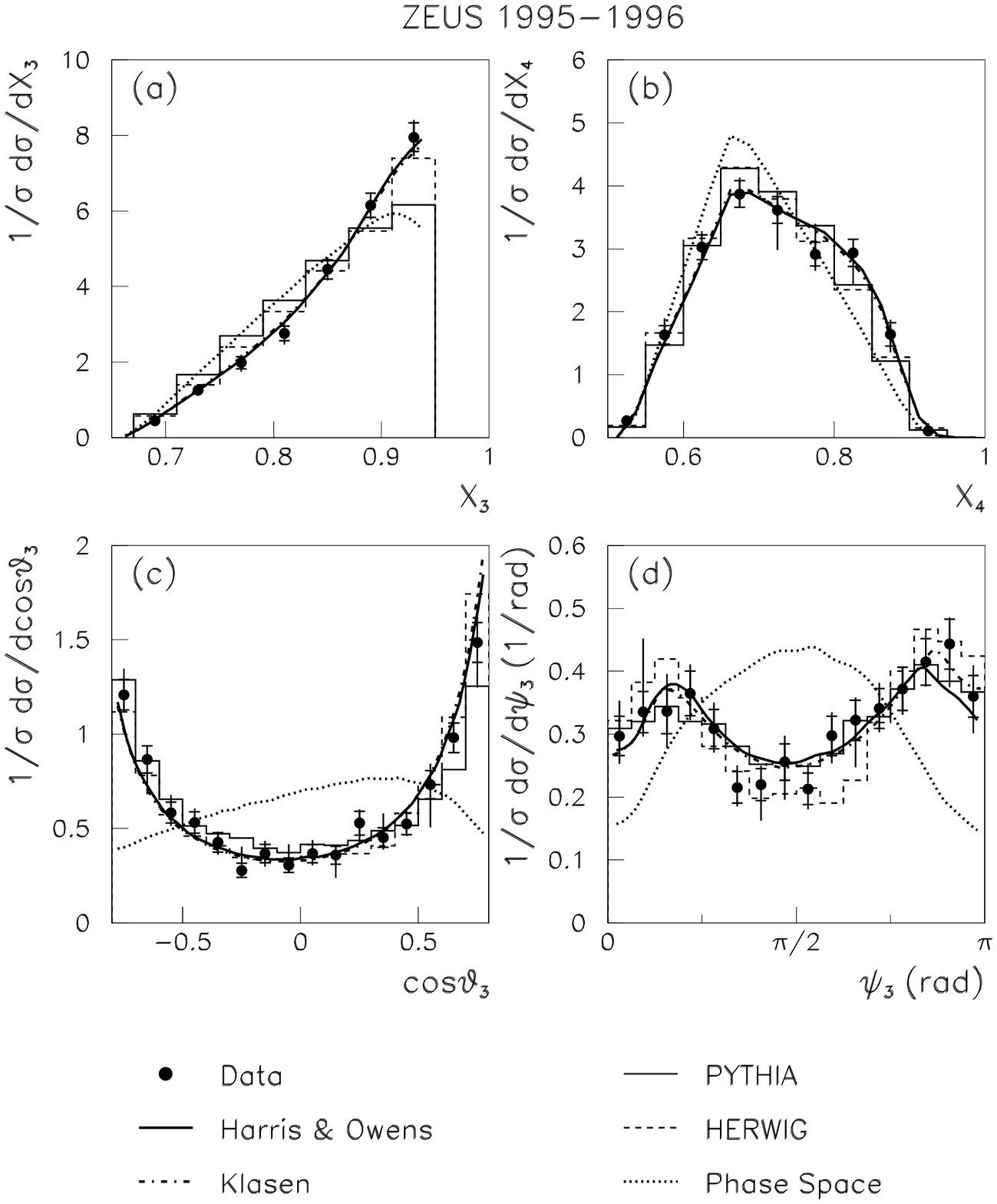,width=11.2cm}}
\put (-0.1,2.2){\epsfig{figure=white.eps,width=5.8cm,height=5.5cm}}
\put (0.0,-3.4){\epsfig{figure=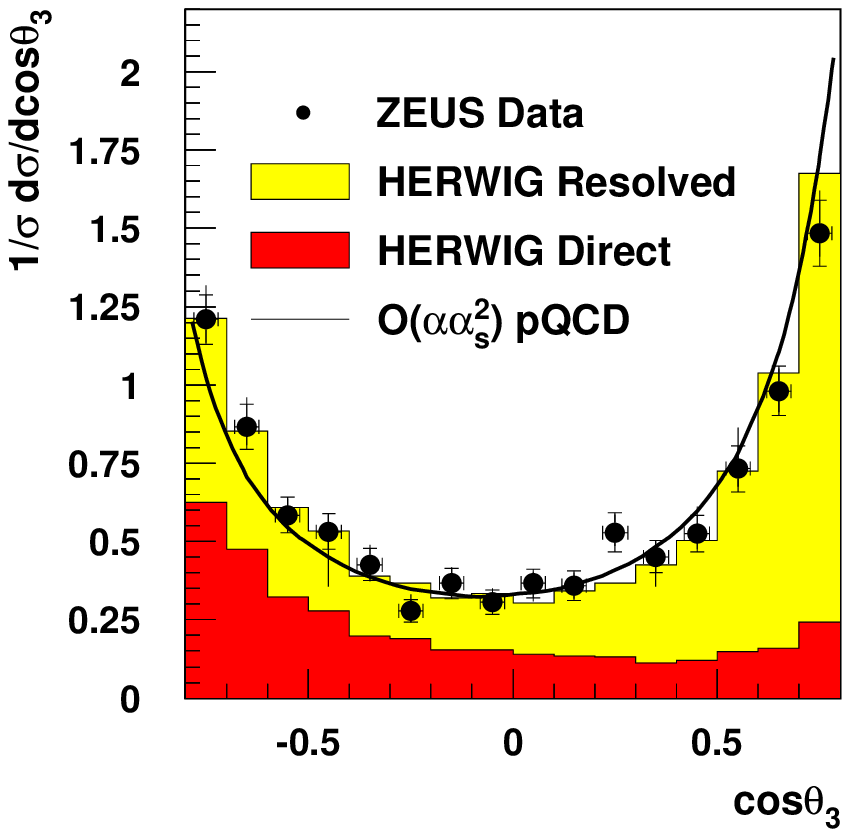,width=12cm}}
\put (1.7,7.0){\small (c)}
\end{picture}
\end{center}
\vspace{-0.5cm}
\caption{\label{fig13} (a) $X_3$ and (b) $X_4$ distributions, and the angular
distributions (c) $\cos\theta_3$ and (d) $\psi_3$.
$\oaa$ and Monte Carlo calculations are shown for comparison.}
\end{figure*}

Figure~\ref{fig13}a and \ref{fig13}b show the $X_3$ and $X_4$ distributions
measured by ZEUS~\cite{zeus3j} for $M_{3j}>50$~GeV, $|\cos\theta_3|<0.8$ and
$X_3<0.95$. Calculations from different models are compared to the data: a
pure phase space calculation does not describe the data. The $\oaa$ QCD
calculations are in good agreement with the data.

The angular distribution of the lowest energy jet in three-jet events is a
distinct probe of the dynamics beyond LO. The measured $\psi_3$ distribution
(figure~\ref{fig13}d) is drastically different from pure phase space and is
in agreement with the $\oaa$ QCD calculations. The
comparison of the data to parton shower models shows that the data favour
color coherence.

The measured $\costh3$ distribution~\cite{zeus3jn}
(figure~\ref{fig13}c) indicates that the highest energy jet tends to go either
forward (proton direction) or towards the rear (photon direction). The
$\oaa$ QCD calculations and the Monte Carlo models
are in good agreement with the data.

The variable $x_{\gamma}^{OBS}$,

\vspace{0.25cm}
\centerline{
$\xo\equiv{\sum_{jets}\etjet e^{-\etajet}\over 2yE_e}$}
\vspace{0.25cm}

\parindent 0em
gives the fraction of the photon energy invested in the production of the
three-jet system and can be used to define resolved and direct processes in a
meaningful way to all orders.
\parindent 0.8em

\begin{figure*}
\begin{center}
\setlength{\unitlength}{1.0cm}
\begin{picture} (10.0,12.8)
\put (-2.0,0.0){\epsfig{figure=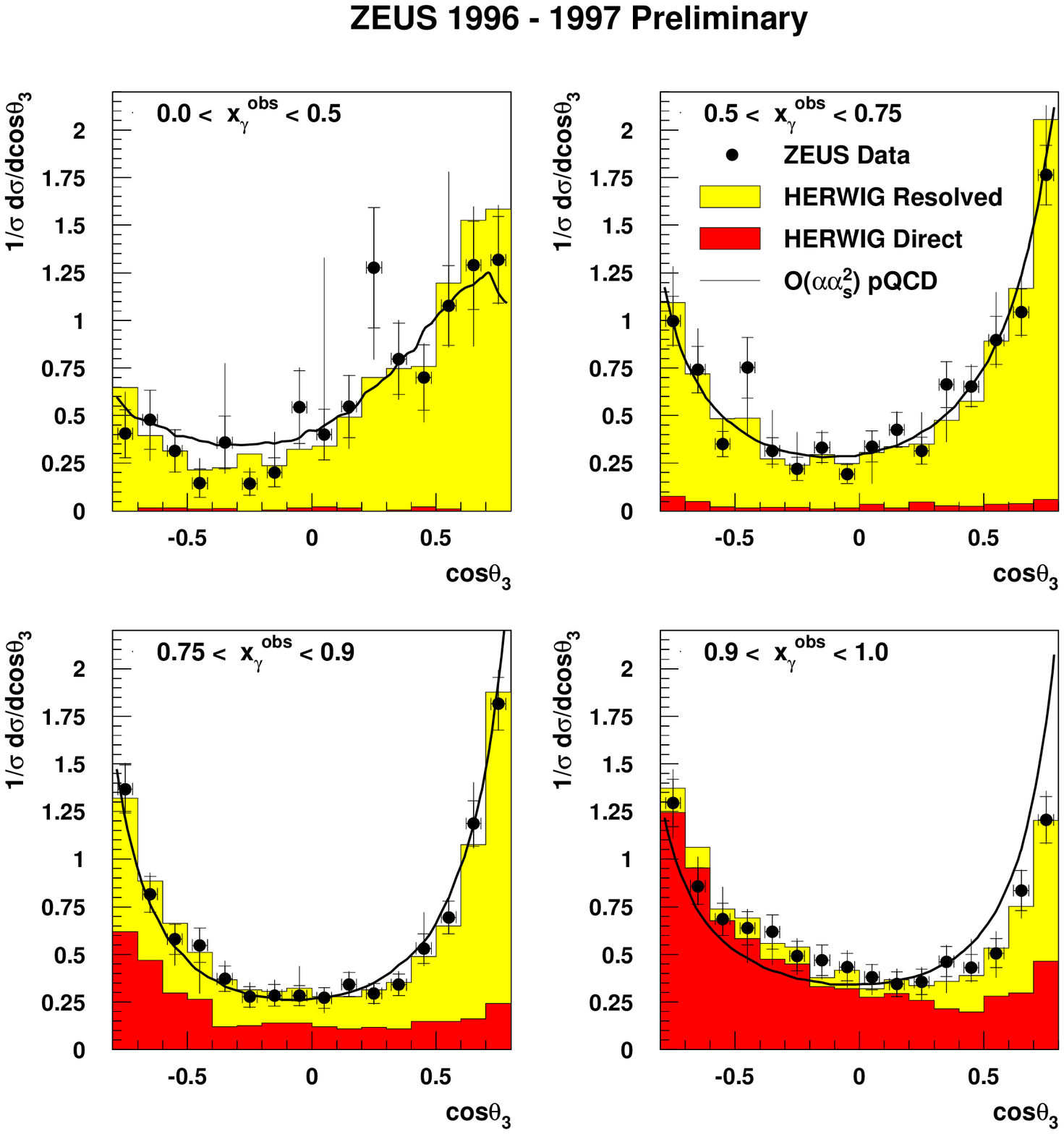,width=13.6cm}}
\end{picture}
\end{center}
\vspace{-0.5cm}
\caption{\label{fig14} $\costh3$ distribution. $\oaa$ and Monte Carlo
calculations are shown for comparison.}
\end{figure*}

Figure~\ref{fig14} shows the $\costh3$ distribution for different regions of
$\xo$. The data indicate that the highest energy jet tends to go in the
forward direction for $\xo<0.5$ (where LO resolved processes dominate) and it
goes more backward as $\xo$ increases. The $\oaa$ QCD calculations are in 
good agreement with the data except for $0.9<\xo<1$.

\section{Conclusions}
Significant progress in comparing measurements and QCD calculations of jet
cross sections in $\gp$ interactions has been achieved; experimental and
theoretical uncertainties have been reduced.

NLO QCD calculations of inclusive jet and dijet cross sections and
$\oaa$ QCD calculations of three-jet cross sections
describe reasonably well the measurements. No significant deviation with
respect to QCD predictions has been observed within the measured range of
the variables studied.

\section*{Acknowledgements}
I would like to thank my colleagues from H1 and ZEUS for their help in
preparing this report.

\end{document}